\documentstyle[11pt]{article}
\def\ref{par\noindent\hangindent=6mm\hangafter=1}
\begin{document}
%\leftline{ll Reunion Iberoamericana de Optica, Guanajuato, Sept. 1995}
%\leftline{
%Invited Talk on Sept. 22, 1995 [S15-1]}
\vbox{
%\rightline{IFUG-95/05 r}
\rightline{\scriptsize{quant-ph/9504018}}
\rightline{\scriptsize{SPIE Vol. 2730, pp. 436-439 (1996)}}
}
%\baselineskip 8mm
%\draft

\begin{center}
{\bf Supersymmetric features of the Maxwell fish-eye lens}

\bigskip

Haret C. Rosu  %$^{a}$\cite{byline},
\\
Instituto de F\'{\i}sica de la Universidad de Guanajuato,
Apdo. Postal E-143, L\'eon, Guanajuato, M\'exico,\\

Marco Reyes %$^{b}$\cite{byline},
\\
Depto. de F\'{\i}sica, CINVESTAV, Apdo. Postal 14-740,
M\'exico D.F., M\'exico,\\

Kurt Bernardo Wolf \\   %$^{c}$\cite{byline},
Instituto de Investigaciones en Matem\'aticas Aplicadas y en
Sistemas-Cuernavaca, Universidad Nacional Aut\'onoma de M\'exico,
Apdo Postal 139-B, 62191 Cuernavaca, Morelos, M\'exico,\\

Octavio Obreg\'on \\   %$^{a,d}$\cite{byline}
Instituto de F\'{\i}sica de la Universidad de Guanajuato,
Apdo. Postal E-143, L\'eon, Guanajuato, M\'exico, and
Universidad Aut\'onoma Metropolitana, Apdo. Postal 55-534, M\'exico
D.F., M\'exico

\end{center}

\bigskip
%\bigskip

\centerline{{\bf ABSTRACT}}
\bigskip
We provide a supersymmetric analysis
of the Maxwell fisheye (MF) wave problem at zero energy.
Working in the so-called $R_{0}=0$ sector, we obtain
the corresponding superpartner (fermionic) MF effective potential
within Witten's one-dimensional (radial) supersymmetric procedure.

\bigskip


\vskip 0.6cm
%\newpage
   %%%%%%%%%%%%%%%%%%%%%%%%     THE PAPER   %%%%%%%%%%%%%%%%%%%%%%%%%%
%%%%%%%%%%%%%%%%%%%%%%%%   written by H.C. Rosu  %%%%%%%%%%%%%%%%%%%%%%%
%%%%%%%%%%%%%%%%%%%%%%%%%%%% March-April 1995 %%%%%%%%%%%%%%%%%%%%%%%%%%%
\bigskip
\centerline{\bf \underline{1. INTRODUCTION}}
\bigskip
Since it is one of the most symmetric systems in the world, the MF lens
has been tackled by many authors.
The well-known analogs of the optical MF
are
the Kepler problem$^{1}$ and the hydrogen atom
$^{1,2,3}$. The classical
mechanical counterpart of the MF spherical waveguide/lens,
$n(r)=\frac{2R^2}{R^2 +r^2},$ $(r\leq R)$, is the motion of a particle with
zero energy (i.e., zero velocity at infinity) in the spherically symmetric
potential $U(r)=-\frac{w}{2R^2(1+r^2/R^2)^2}$.
In this paper we shall consider the MF wave/quantum problem at
fixed null energy, (or zero-binding energy),
$[-\frac{\hbar ^2}{2m}
\nabla ^2_r + U_{MF}(r)]\psi ({\bf r})=0$, with
$U_{MF}(r)=
-\frac{w{\cal E}_0}{[1+(r/R)^2]^2}$,
where $w>0$ is a coupling constant, and $R>0$ is the radius of the lens.
An energy scale ${\cal E}_0=\hbar ^2/2mR^2$ has been introduced for the
potential part$^{4}$.

Demkov and Ostrovsky included the MF wave problem in a more
general type of focusing
potentials$^{1}$ bearing their name$^{5}$ and characterized by a parameter
$\kappa$, which is unity for the fisheye.
They have shown
that for the cases $\kappa =k_1/k_2$, with $k_1$ and $k_2$ integers,
i) the classical
trajectories of a zero-energy (i.e., zero velocity at infinity) particle
close after $k_2$ revolutions around the force centre, and ii) all the
trajectories passing through a given point come to a focus after $k_2/2$
revolutions.

The above wave equation can be written in terms of
the scaled variable $\rho =r/R$, (hereafter the energy scale is to be
understood), as follows

$$[-\frac{\partial ^2}{\partial \rho ^2}-\frac{2}{\rho}
\frac{\partial}{\partial \rho}+ \frac{l(l+1)}{\rho ^2} -
\frac{w}
{(1+\rho ^{2})^2}]\psi ({\bf \rho})=0.   \eqno(1)$$
It is quite straightforward to solve the
Sturm-Liouville problem, Eq.~(1). Moreover, it can be turned into an eigenvalue
problem for the coupling constant, $w$, and can also be written as a Laplace
equation on the four-dimensional sphere$^{1}$.
The known results are the following.
When $w$ assumes the quantized values,
$w_{n}=4n^2-1$, where $n$ is the MF principal quantum number (see below),
the regular, normalizable (``bound") solutions, decreasing at infinity, read
$$\psi _{nlm}({\bf \rho})=R_
{nl}(\rho)Y_{lm}(\theta, \phi)=
\frac{{\cal N} _{nl}}
{\rho ^{-l}(1+\rho ^{2})^{(2l+1)/2}}
C^{l+1}_{n-1-l}(\xi)Y_{lm}(\theta, \phi),   \eqno(2)$$
where $\xi=\frac{1-\rho ^{2}}{1+\rho ^{2}}$,
$n=n_r +l+1$, $n_r =0,1,2,...$, are the MF
principal and radial quantum numbers, respectively, $l$ and $m$ are the
spherical harmonic numbers, $C_{p}^{q}(\xi)$ are the
Gegenbauer polynomials,
i.e., the solutions of the corresponding ultraspherical equation (see Eq.~(5)
below), and ${\cal N}_{nl}$ are the normalization constants.
What one gets when the $w$ parameter becomes larger and larger is an increase
of the degeneracy of the ``bound'', E=0 state, but only for the quantized
values $w_{n}$.
The degree of degeneracy of such a group of states is $n^2$,
($n$ = 1,2,3...), similar to the electron energy levels in a Coulomb field.
%%%%%%%%%%%%%%%%%%%%%%%%%%%%%%%%%%%%%%%%%%%%%%%%%%%%%%%%%%%%%%%%%%%%%%%%%%
\bigskip

\centerline{\underline{\bf 2. MF SUPERSYMMETRY IN THE $R_0$ SECTOR}}
\bigskip

Since we want to apply the supersymmetric Natanzon-like scheme
as discussed for example by L\'evai$^{6}$, we pass to
the functions $u_{nl}=\rho R_{nl}$ fulfilling the one-dimensional radial
equation

$$[-\frac{\partial ^2}{\partial \rho ^2} + U_{eff}^{-}(\rho)]u\equiv
[-\frac{\partial ^2}{\partial \rho ^2} + \frac{l(l+1)}{\rho ^2}-
\frac{4n^2-1}
{(1+\rho ^{2})^2}]u\equiv H^{-}u=0,
\eqno(3)$$
in which we have already included supersymmetric superscripts.
The functions $u_{nl}$ are of the type
$f_l(\rho)C_{n-1-l}^{l+1}(\xi (\rho))$, where $f_l(\rho)$
reads
$$f_l(\rho)=\frac{\rho ^{l+1}}{(1+\rho ^{2})^{(2l+1)/2}}, \eqno(4)$$
and
the Gegenbauer polynomials, $C_{p}^{q}$, of degree $p=n_r$ and
parameter $q=l+1$,
are the solutions of a
second-order differential (ultraspherical) equation of the type

$$P(\xi)\frac{d^2 C}{d\xi ^2}+Q_{l}(\xi)\frac{dC}{d\xi} +R_{p}(\xi)C(\xi)= 0,
\eqno(5)$$
with
$P(\xi)=1$, $Q_{l}(\xi)=\frac{(2l+3)\xi}{\xi ^2 -1}$
and
$R_{p}(\xi)=-\frac{p(2q+p)}{\xi ^2 -1}$,
in which we emphasized the indexing of the $R$- functions according to
the various sectors $p=n_r$ (0,1,2,...), which is a quite general
characteristic of orthogonal polynomials$^{6}$.

In the Natanzon-like scheme, the following equations can be obtained
$$\frac{\xi ^{''}}{(\xi ^{'})^2} + \frac{2f_l^{'}}{\xi ^{'}f_l}=Q_l(\xi(\rho)),
\eqno(6)$$
and
$$\frac{f_l^{''}}{(\xi ^{'})^2 f_l}- \frac{U_{eff}^-}{(\xi ^{'})^2}=
R_p(\xi (\rho)),
\eqno(7)$$
where $U_{eff}^-$ is given in Eq.~(3).
From Eq.~(6) the function $f_l(\rho)$ can be written as follows

$$f_l(\rho)\sim (\xi ^{'}(\rho))^{-1/2}\exp[\frac{1}{2}
\int ^{\xi(\rho)}Q_l(\xi(\rho))d\xi].  \eqno(8)$$

One can then define the `ground state'
by means of the $R_{0}(\xi)=0$ sector, within which the Gegenbauer
polynomials are $C_{0}^{q}=1$ for any $q$.
In this simple case, from Eq.~(7) one gets

$$U_{eff}^-=+W_l^2(\rho) -\frac{d W_l}{d\rho},  \eqno(9)$$
with $W_l(\rho)=-\frac{d}{d\rho}\ln f_l(\rho)$. Eq.~(9) is
the standard Riccati equation in ordinary (Witten) 1D supersymmetric
quantum mechanics$^{7}$.
Since we actually know from Eq.~(4) the function $f_l(\rho)$ in the MF case
(one can check that Eq.~(8) leads to the same function), a
short calculation gives the MF superpotential
$$W_{MF}(\rho)=\frac{l}{\rho}-\frac{2l+1}{\rho(1+\rho ^{2})}.
\eqno(10)$$
The effective MF superpartner in the $R_{0}=0$ sector is obtained by changing
the sign of the derivative in the Riccati equation
%$$U_{eff}^{-}=-\frac{dW_{MF}(\rho)}{d\rho}+W_{MF}^{2}(\rho),
%\eqno(14a)$$

$$U_{eff}^{+}=+\frac{dW_{MF}(\rho)}{d\rho}+W_{MF}^{2}(\rho).
\eqno(11)$$
Thus,
$$U_{eff}^{-}=\frac{l(l+1)}{\rho ^2}-
\frac{(2l+1)(2l+3)}{ (1+\rho ^{2})^2},
\eqno(12)$$
and
$$U_{eff}^{+}=\frac{l(l-1)}{\rho ^2}-
\frac{(2l+1)(2l-3)}
{(1+\rho ^{2})^2}
+\frac{2(2l+1)}{\rho ^{2}(1+\rho ^{2})^2}.
\eqno(13)$$
We have plotted $U_{eff}^{-}$ and $U_{eff}^{+}$ for some values of the
parameters in Fig.~1.
The MF factorization operators
$A_{MF}=\frac{d}{d\rho} + W_{MF},$
and
$A^{+}_{MF}=-\frac{d}{d\rho}+ W_{MF}$ can be used to write the MF fermionic
equation.
From the plot of the MF fermionic potentials one can notice their
repulsive nature. In the supersymmetric sense this means the dissapearance
of the zero-energy `ground state' for the superpartner Hamiltonian.
Consequently, the fermionic equation
should be written in the continuum
$$H^{+}u_{1}\equiv AA^{+}u_{1}\equiv
(-\frac{d^2}{d\rho ^{2}}+U_{eff}^{+})u_{1}=
k^2 u_{1},\;\; k \in (0,\infty).   \eqno(14)$$
This equation will be studied elsewhere. Here we remark that in order
to get the supersymmetric increment in the effective potential
we used only the particular solution of the Riccati equation, Eq.~(9).
On the other hand, the connection with the Gel'fand-Levitan
inverse scattering method requires the general Riccati solution$^{8}$,
which we have worked out recently$^9$.

In conclusion, we have presented the supersymmetric structure
of the $R_{0}=0$ sector of the MF problem.
In this way, we were able to introduce the MF fermionic effective
potentials. Moreover, we have found numerically their trapping region
(see Fig. 1b).

\bigskip
%%%%%%%%%%%%%%%%%%%%%%%%%%%%%%%%%%%%%%%%%%%%%%%%%%%%%%%%%%%%%%%%%%%%%
\centerline{\underline{\bf 3.~ACKNOWLEDGEMENTS}}
\bigskip

This work was partially supported by the CONACyT Projects 4862-E9406,
4868-E9406, DGAPA IN 1042 93 at UNAM-M\'exico, and a CONACyT Graduate
Fellowship.

\bigskip
\newpage
%%%%%%%%%%%%%%%%%%%%%%%%%%%%%%%%%%%%%%%%%%%%%%%%%%%%%%%%%%%%%%%%%%%%%%
\centerline{\underline{\bf 4.~REFERENCES}}
\bigskip
%\begin{thebibliography} {99}

%\bibitem[*]{byline} E-mail: rosu@ifug.ugto.mx; corresponding author
%\bibitem[*]{byline} E-mail: mareyes@fnalv.fnal.gov
%\bibitem[*]{byline} E-mail: bwolf@redvax1.dgsca.unam.mx
%\bibitem[*]{byline} E-mail: octavio@ifug.ugto.mx

%\bibitem{do}
1. Yu.N. Demkov and V.N. Ostrovsky, ``Internal symmetry of the
Maxwell ``fish-eye" problem and the Fock group for the hydrogen atom",
Sov. Phys. JETP vol. 33, pp. 1083-1087, 1971.

%\bibitem{b}
2. H.A. Buchdahl, ``Kepler problem and Maxwell fish-eye", Am. J. Phys.
vol. 46, pp. 840-843, 1978.

%\bibitem{ch}
3. A.C. Chen, ``The Coulomb problem and the Maxwell fish-eye problem",
Am. J. Phys. vol. 46, pp. 214-217, 1979; ``The zero-energy Coulomb problem",
J. Math. Phys. vol. 19, pp. 1037-1040, 1978.

%\bibitem{dn}
4. J. Daboul and M.M. Nieto, ``Quantum bound states with zero binding
energy", Phys. Lett. A vol. 190, pp. 357-362, 1994.

%\bibitem{kb}
5. Y. Kitagawara and A.O. Barut, ``Period doubling in the n+l filling
rule and dynamical symmetry of the Demkov-Ostrovsky atomic model",
J. Phys. B: At. Mol. Phys. vol. 16, pp. 3305-3327, 1983; ``On the dynamical
symmetry of the periodic table: ll. Modified Demkov-Ostrovsky atomic model",
ibid. vol. 17, pp. 4251-4259, 1984.

%\bibitem{lev}
6. G. L\'evai, ``A search for shape-invariant solvable potentials",
J. Phys. A: Math. Gen. vol. 22, pp. 689-702, 1989.

%\bibitem{w}
7. E. Witten, ``Dynamical breaking of supersymmetry", Nucl. Phys. B vol. 185,
pp. 513-554, 1981.

%\bibitem{n}
8. M.M. Nieto, ``Relationship between supersymmetry and the inverse method
in quantum mechanics", Phys. Lett. B vol. 145, pp. 208-210, 1984.

9. H.C. Rosu, M. Reyes, K.B. Wolf, and O. Obreg\'on, ``Second solution of
Demkov-Ostrovsky superpotentials", Los Alamos electronic preprint archive
quant-ph/9505006, 1995.

%\end{thebibliography}
%%%%%%%%%%%%%%%%%%%%%%%%%%%%%%%%%%%%%%%%%%%%%%%%%%%%%%%%%%%%%%%%%%%%%%%%%

%\newpage
%{\bf Figure Captions}
\bigskip

%Fig. 1.  The effective DO superpartners at zero energy,
%$U_{eff}^{\mp}(\rho)$, (a) and (b),
%respectively, in units of ${\cal E}_{0}$, for  $l=2$, and $\kappa$ =
%1/2, 1, and 3/2.

\bigskip
\bigskip
\bigskip
\bigskip
\bigskip
\bigskip
\bigskip
\bigskip
\bigskip
\bigskip
\bigskip
\bigskip
\bigskip
\bigskip
\bigskip
\bigskip
\bigskip
\bigskip

\centerline{\bf Fig. 1.}

MF superpartners of the $R_0=0$ sector in ${\cal E} _0$ units:
(a), $U^{-}_{eff}$ for $l$= 1, 5, 10,
and (b), $U^{+}_{eff}$ for
$l$=6, 7, 8. We have
plotted $U^{+}_{eff}$ in the region of the critical (inflexion) angular
number, $l_{cr}$, that
we have found numerically to be $l_{cr}$=6.876 for $\rho _{cr}$=1.599. The
critical $l$ is the entry point toward a pocket (trapping) region of
$U^{+}_{eff}$ for $l>l_{cr}$.

%%%%%%%%%%%%%%%%%%%%%%%%%%%%%%%%%%%%%%%%%%%%%%%%%%%%%%%%%%%%%%%%%%%%%%%%%%

%The principal q. number $N$ labels nonrel. Coulomb energy levels through
%the
%Balmer formula. The radial q. number is the number of nodes of the radial
%wave number between 0 and $\infty$ plus one, $n_r=n_n+1. We have also
%$N=n_r+l$. (A.K. Grant et al., hep-ph/9506315, PR D

\end{document}